\documentclass[useAMS,usenatbib]{mn2e}

\usepackage[english]{babel}
\usepackage[latin1]{inputenc}
\usepackage{amssymb}
\usepackage{graphicx}
\usepackage{color}
\usepackage{latexsym}
\usepackage{multirow}

\title[Automated SDSS physical classification]{Automated physical classification in the SDSS DR10. A catalogue of candidate Quasars.}
\author[M. Brescia, S. Cavuoti \& G. Longo]{M. Brescia$^{1}$\thanks{E-mail:
brescia@na.astro.it}, S. Cavuoti$^{1}$, G. Longo$^{2}$\\
$^{1}$Astronomical Observatory of Capodimonte - INAF, via Moiariello 16, I-80131, Napoli, Italy\\
$^{2}$Department of Physics, University Federico II, Via Cinthia 6, 80126 Napoli, Italy}
\begin{document}

\date{Accepted, 13 April 2015. Received, 7 April 2015; in original form, 20 November 2014}

\pagerange{\pageref{firstpage}--\pageref{lastpage}} \pubyear{2015}

\maketitle

\label{firstpage}

\begin{abstract}
We discuss whether modern machine learning methods can be used to characterize the physical nature of the large number of objects
sampled by the modern multi-band digital surveys.
In particular, we applied the MLPQNA (Multi Layer Perceptron with Quasi Newton Algorithm) method to the optical data of the Sloan Digital Sky Survey - Data Release 10,
investigating whether photometric data alone suffice to disentangle different classes of objects as they are defined in the SDSS spectroscopic classification.
We discuss three groups of classification problems:
\textit{(i)} the simultaneous classification of galaxies, quasars and stars;
\textit{(ii)} the separation of stars from quasars;
\textit{(iii)} the separation of galaxies with normal spectral energy distribution from those with peculiar spectra, such as starburst or starforming galaxies and AGN.
While confirming the difficulty of disentangling AGN from normal galaxies on a photometric basis only, MLPQNA proved to be quite effective in the three-class separation.
In disentangling quasars from stars and galaxies, our method achieved an overall efficiency of $91.31\%$ and a QSO class purity of $\sim95\%$. The resulting catalogue of candidate quasars/AGNs consists of $\sim 3.6$ million objects, of which about half a million are also flagged as robust candidates, and will be made available on CDS VizieR facility.
\end{abstract}

\section{Introduction}\label{SEC:introduction}
Broad band photometry from wide-field imagers mounted on dedicated telescopes and instruments has been and will continue to be our main source of information for a large fraction of the extragalactic universe. Spectroscopy provides a more detailed and deeper understanding of the physical properties of individual objects than photometry. However, spectroscopy will never be able to fully sample the populations of galactic or extragalactic objects in either number or depth.
It is therefore of great interest to determine whether it is possible to extract estimates of physical parameters, such as distance, metallicity, star formation rate, morphology and the presence or absence of an active nucleus, from the coarse information provided by photometry.

For a given object, photometric quantities, such as magnitudes in different bands, momenta of the light distribution and morphological indexes, define its position in a high dimensional parameter space which we shall call the Observed Parameter Space (or OPS, cf. \citealt{george12}).
The de-projection of the OPS into the Physical Parameter Space (PPS, i.e. the parameter space defined by the physical quantities), is however a complex operation, made in some cases almost impossible by the degeneracy existing in both the data and the physical parameters themselves.
Most of the time, such de-projections require an intermediate passage, i.e. the de-projection of the OPS onto the Spectroscopic Parameter Space (or SPS), i.e. the space defined by observable spectroscopic quantities such as redshift, equivalent widths of specific spectral lines and spectroscopic indexes.
The parameters defining the SPS are in fact more directly related to the intrinsic physical properties of the objects.
Spectroscopy, for instance, is more effective than photometry in disentangling normal galaxies from those hosting an AGN, or in classifying different types of AGN in broad classes (e.g., Seyfert, LINERS, etc.).
Actually, the definition of nearly all AGN classes is based on spectroscopic criteria through the equivalent widths of some lines (see for instance \citealt{kewley2001,lamareille2010}).

The problem of mapping the OPS onto some subspaces of the SPS and then on the PPS, is frequently encountered in the literature.
Examples include the determination of photometric redshifts, which is an attempt to reproduce what is decidedly a spectroscopic quantity (the redshift) using photometry alone; the search for candidate quasars and blazars using photometric diagnostics; or the search for candidate AGN and starburst galaxies.

Usually these mapping functions are quite complex and difficult (if not plainly impossible) to be derived in simple analytical forms and require statistical or empirical approaches which are common in the field of Machine Learning (hereinafter ML).
In fact, if higher accuracy information is available for a subset of objects (in the case discussed here, spectroscopic parameters or flags in the SPS), it can be used to teach a ML method how to map the available data (the photometric data in the OPS) into the higher accuracy ones.
In the ML field this approach is usually called supervised learning. We emphasize that, while supervised learning is very powerful in uncovering the hidden relationship between input parameters and the so-called knowledge base (hereafter KB), there are two main limitations:

\begin{itemize}
\item the mapping function cannot be extrapolated outside of the region of the OPS that is properly sampled by the SPS;
\item any bias present in the KB is necessarily reproduced (if not amplified) in the output.
\end{itemize}

To better exemplify these two points let us focus on the main topic addressed in this work where we use a specific ML method, the MLPQNA \citep{brescia2013,cavuoti2014}, to tackle a particular incarnation of the so called \textit{Physical classification of galaxies} problem. We will specifically address the possibility of disentangling normal, non-active galaxies from those hosting an AGN using only photometric measurements.

In a previous paper we have discussed the use of the same method to classify different types of AGN \citep{cavuoti2014}.
In this work we shall instead focus on the question of whether it is possible to use only optical colors to produce reliable (or at least statistically well controlled) AGN candidates and to disentangle normal-inactive galaxies from those hosting an AGN.

As already pointed out in \cite{cavuoti2014}, in spite of the unique physical mechanism responsible for the nuclear activity \citep{antonucci1993,padovani1995}, the phenomenological complexity of AGNs is so high that there cannot be a unique method equally effective in identifying all AGN phenomenologies in every redshift range (cf. \citealt{messias2010}).
Moreover, AGN and starbursts, long studied separately, are now thought to be correlated \citep{Schweizer2006,pilbratt2010} and difficult to disentangle without using mid-IR data.

Furthermore, we also investigate the possibility of identifying a catalogue of candidate QSOs, by comparing the efficiency of classification in terms of a compromise between purity and completeness following two different strategies: i) by exploiting a self-consistent strategy starting from three different classes of objects (for instance galaxies, stars and QSOs); ii) by assuming a pre-determined separation between resolved and un-resolved objects (star/galaxy), we performed the usual two-class classification between stars and QSOs.

The paper is structured as follows. In Section~\ref{SEC:thedata} we discuss in same detail the data. In Section~\ref{SEC:themethod} we summarize the general methodology adopted and provide a short description of the ML method used. In Section~\ref{SEC:experiments} we describe in some detail the various classification experiments performed and, finally, in Section~\ref{SEC:conclusions} we discuss the results.

\section{The data}\label{SEC:thedata}
Among the countless merits of the Sloan Digital Sky Survey \citep{york2000,stoughton2002}, there is also the fact that it has paved the way to the experimenting and wide adoption within the astronomical community of innovative methods based on ML (or on statistical pattern recognition), thus fostering the birth of the emerging field of Astroinformatics \citep{borne2010}.

The SDSS in fact provides an ideal test ground for ML algorithms and methods: a large and complex photometric database with hundreds of features measured for hundreds of millions of objects (defining the SDSS subregion of the OPS), complemented  by spectroscopic information for a significant subsample (roughly $\sim 1\%$) of objects (i.e. the SPS).
Furthermore, it offered to different groups, working with different methodologies, the possibility to address significant problems, thus allowing a robust and fair assessment of the performance of any new method.

Even constraining ourselves to the specific interest of the present work, it would be difficult to summarize the applications of ML methods to SDSS data and we shall just mention two.
The search for candidate quasars has been widely discussed in \citealt{richards2004,richards2009,dabrusco2009,abraham2012}.
while the characterization of AGN has been discussed in \cite{cavuoti2014}.
Last but not least, ML methods have been crucial for the first successful astronomical citizen science project: the galaxy Zoo \citep{lintott2008}, where empirical methods were used to compare and evaluate the performance of a large number of human simple, repetitive, and independent classification tasks.

In this paper we use photometric and spectroscopic data extracted from the SDSS Data Release 10 (DR10; \citealt{ahn2013,eisenstein2011}).
The DR10 photometry covers $14,555$ $deg^2$ of the celestial sphere for a total of more than $469$ millions unique (i.e. without duplicates, overlaps and repeated measurements) objects, but not necessarily unique \textit{astrophysical} objects. DR10 also includes spectroscopic information for more than $3$ million objects.
These data come from a wide range of concurrent experiments: the $7$ initial SDSS data releases, the Sloan Extension for Galactic Understanding and Exploration (SEGUE-2, \citealt{yanny2009}), the Apache Point Observatory Galaxy Evolution Experiment (APOGEE; \citealt{apogee}) mainly targeted to Milky Way stars; the Barion Oscillation Spectroscopic Survey (BOSS; \citealt{boss}) and, finally, the Multi Object APO Radial Velocity Exoplanet Large-Area Survey (MARVELS; \citealt{marvels}).

In  order to build our KB we used the spectroscopic classification flags provided by the SDSS DR10. It includes also a spectroscopic classification (parameters $CLASS$ and $SUBCLASS$) obtained by the SDSS team through the comparison of individual spectra with templates and from equivalent width ratios. This classification is  articulated in three main classes: $GALAXY$, $QSO$ and $STAR$ as follows:\\

\noindent \textbf{\textit{GALAXY}}. Objects having a galaxy template. Its subclasses are:

\begin{itemize}
\item  STARFORMING: based on whether the galaxy has detectable emission lines that are consistent with star-formation according to the criteria:
$$log10\left(OIII/H_{\beta}\right)<0.7- 1.2\left(log10\left(NII/H_{\alpha}\right)+0.4\right)$$
\item STARBURST: set if the galaxy is star-forming but has an equivalent width of $H_{\alpha}$ greater than $50$ \AA;
 \item AGN: based on whether the galaxy has detectable emission lines that are consistent with being a Seyfert or LINER according to the relation:
$$log10\left(OIII/H_{\beta}\right)>0.7- 1.2\left(log10\left(NII/H_{\alpha}\right)+0.4\right)$$
\end{itemize}

\noindent \textbf{\textit{QSO}}. Objects identified with a QSO template. Based on the SDSS definition, if a galaxy or quasar has spectral lines detected at the $10$ sigma level with $\sigma > 200 km/sec$ at the $5$ sigma level, the indication \textit{BROADLINE} is appended to their subclass.\\

\noindent \textbf{\textit{STAR}}. Objects matching a stellar template (among $26$ spectral subclasses). Since we were not interested in these subclasses, they will
be ignored in what follows.\\

The total number of spectroscopic objects is $3,079,151$ divided in different types as summarized in Table \ref{TAB:KBdetails}.
We assumed all galaxies not flagged otherwise to be normal, i.e. neither AGN or starburst.

It is worth noting that sharp cuts based on crisp thresholds introduce ambiguities in the mapping of the SPS onto the PPS.
To be more clear: a large fraction of the objects with spectroscopic features near the threshold values may be misclassified (i.e a galaxy hosting an AGN may be erroneously put in the normal galaxy bins due spectroscopic errors and viceversa). These ambiguities are intrinsic to the KB and will affect the projection of the OPS onto the SPS and therefore the PPS.

For all spectroscopic objects we downloaded the sets of magnitudes listed in the upper part of Table \ref{TAB:features}.
The choice of using different sets of magnitudes was dictated by the fact that, since we were interested also in finding AGN candidates, different apertures can be used to weight in different ways the contribution of the central unresolved source and of the extended surrounding galaxy.

\begin{table*}
\centering
\begin{tabular}{l|l|r|r}
\hline
{\bf CLASS}              & {\bf SUBCLASS}              &{\bf Nr. of obj. }     & {\bf Total Nr.}\\
\hline
\hline
                         & AGN                         &      22,589           &           \\
                         & BROADLINE                   &      18,629           &           \\
                         & STARBURST                   &      73,166           &           \\
                GALAXY   & STARFORMING                 &     265,704           &   385,298 \\
                         & AGN BROADLINE               &       3,505           &           \\
                         & STARBURST BROADLINE         &         166           &           \\
                         & STARFORMING BROADLINE       &       1,539           &           \\
                         & NORMAL                      &                       & 1,526,215 \\
                         &                             &     sub-total         & 1,911,513 \\
\hline
                         & AGN                         &       970             & \\
                         & BROADLINE                   &   257,572             & \\
                         & AGN BROADLINE               &     2,954             &\\
                QSO      & STARBURST BROADLINE         &     9,127             & 271,603\\
                         & STARFORMING BROADLINE       &       552             & \\
                         & STARBURST                   &       315             &\\
                         & STARFORMING                 &       113             &\\
                         & NORMAL                      &                       & 97,695\\
                         &                             &     sub-total         & 369,298\\
\hline
 STAR                    &                             &                       & 798,340\\
\hline
\end{tabular}
\caption{Partition of the SDSS-DR10 spectroscopic objects into spectroscopic classes and subclasses.
Column 1: gives the broad spectral type; column 2: spectral subtype; column 3: number of objects belonging to that specific subclass;
column 4: Total number of objects in a given class.} \label{TAB:KBdetails}
\end{table*}

\begin{table*}
\centering
\begin{tabular}{l|l}
\hline
{\bf type}   & {\bf parameters}\\
\hline
\hline
Identification          & objID, specObjID\\
Coordinates             & RA, DEC\\
psfMag                  & ugriz magnitudes and related errors $mag\_err$\\
fiberMag                & ugriz magnitudes and related errors $mag\_err$\\
modelMag                & ugriz magnitudes and related errors $mag\_err$\\
cmodelMag               & ugriz magnitudes and related errors $mag\_err$\\
deredMag                & ugriz magnitudes\\
extinction              & ugriz extinction values\\
\hline
spectroscopic redshift  & z, zWarning\\
classification          & type, class, subclass, flags\\
\hline
\end{tabular}
\caption{Description of the parameters (features and targets) used in this work. The first part of the table lists the photometric parameters in the OPS, the second one the spectroscopic data and targets. Column 1: collective name of a given set of parameter; column 2: the corresponding SDSS parameters.} \label{TAB:features}
\end{table*}

We performed a preliminary photometric filtering at the query time, by using the \textit{primary} mode and by taking into account the DR10 prescriptions encapsulated by the flags \textit{calibStatus} and \textit{clean}. Besides these flags, any use of the produced candidate QSOs should be carefully investigated in terms of their photometric reliability \citep{palanque2013}.

Finally, the catalogue had to be cleaned for an handful of objects which appear to be duplicated in the main SDSS-DR10 archive.

\section{The method}\label{SEC:themethod}
Supervised machine learning methods used for classification tasks require an extensive KB formed by objects for which the outcome of the classification (i.e. the target) is a-priori known. The methods learn from these examples the unknown and often complex rule linking the input data (in this case the photometric parameters) to the target (in this case a physical class known from the KB). Since these methods cannot be used for extrapolating knowledge outside the KB ranges, it is crucial to keep in mind that the regions of the OPS adequately sampled by the KB define also the domain of applicability of the method.

The KB is usually split in three different subsets to be used for training, validation and test, respectively.
The training set is used by the method to learn the mapping function, the validation set is mainly used to avoid the so called \textit{overfitting}, i.e. the loss of generalization capabilities, and a blind test set is used to evaluate the performance of the method, by exposing the trained network to objects in the KB which have never been seen before by the network itself.

Evaluation of the results with the a-priori known values of the objects in the blind test set allows evaluation of a series of statistical indicators and parametrization of the performance of the method itself.

An alternative approach, which is also the one used in this work, is the so called \textit{leave-one-out $k$-fold cross validation} which can be implicitly performed during training \citep{geisser1975}. The automatized process of the cross-validation consists of performing $k$ different training runs with the following procedure: (\textit{i}) random splitting of the training set into $k$ random subsets, each one composed by the same percentage of the data set (depending on the $k$ choice); (\textit{ii}) at each run the remaining part of the data set is used for training and the excluded fraction for validation. While avoiding overfitting, the k-fold cross validation leads to an increase of the execution time of $\sim k-1$ times the total number of runs.

In all experiments listed in the following section we used the MLPQNA method and a $5-fold$ cross-validation.

\subsection{The model MLPQNA}\label{SEC:mlpqna}

DAMEWARE (DAta Mining \& Exploration Web Application REsource; \citealt{brescia2014}) is an infrastructure which offers to anyone the possibility of engaging in complex data mining tasks by means of a web-based approach to a variety of data mining methods.
Among the methods available, for the present work we used a Multi Layer Perceptron (MLP; \citealt{rosenblatt1961}) neural network which is among the most used feed-forward neural networks in a large variety of scientific and social contexts.
MLPs may differ widely in both architecture and learning rules. In this work, we used the MLPQNA model, i.e. a MLP implementation (Fig.~\ref{fig:mlpqna}) where the learning rule is based on the Quasi Newton Algorithm (QNA).\\
The QNA relies on Newton's method to find the stationary (i.e. the zero gradient) point of a function and the QNA algorithm is an optimization of the basic Newton learning rule based on a statistical approximation of the Hessian of the training error obtained through a cyclic gradient calculation.
MLPQNA makes use of the well known L-BFGS algorithm (Limited memory - Broyden Fletcher Goldfarb Shanno; \citealt{byrd1994}) which was originally designed for problems with a very large number of features (hundreds to thousands).\\
The MLPQNA method has been extensively discussed elsewhere in the contexts of both classification \citep{brescia2012} and regression \citep{brescia2013,cavuoti2015,cavuoti2012,cavuoti2014b}; a first attempt to use MLPQNA for a similar problem, namely the classification of emission line galaxies, was presented in \cite{cavuoti2014}.

\section{Experiments}\label{SEC:experiments}
We now describe the main classification experiments which were performed. Since the general astronomer may not be familiar with the ML methodology and could be confused in dealing with the following sections, we wish to emphasize a few points.

Most of ML methods are not deterministic and there is no clear-cut rule to optimize the parameters of the models. This is particularly true in the astronomical case, due to classification degeneracy in some parts of the OPS. Thus it is often not possible to establish a-priori which combination of input parameters is optimal for a given task.
This implies that, in order to find the optimal method, many experiments with different settings need to be run and evaluated. A full understanding of the performance of a specific experiment can be achieved only through the comparison of the outcomes of a large number of experiments, each run with different settings.

Our experiments can be divided into three main families:
\begin{itemize}
\item two-class experiments for disentangling normal galaxies from other types (normal - AGN/QSO).
\item two-class experiment to disentangle quasars from stars (QSO - STAR);
\item three-class experiments (STAR - GALAXY - QSO);
\end{itemize}

Before entering into a detailed description of these three groups of experiments, it is useful to list the criteria which were adopted in order to reduce the number of experiments.\\

\textit{Input features and targets}.
As input features we used the $5$ $psfMag$ and the $5$ $modelMag$ magnitudes provided by the SDSS in the $u,g,r,i,z$ photometric system. As target vectors we used the spectroscopic taxonomy provided by SDSS and discussed in Sec.~\ref{SEC:thedata}.
Following a similar approach but using a different ML method, \cite{abraham2012} found that better performance can be obtained by using $10$ colors, obtained from all possible combinations of the SDSS $psfMag$ magnitudes, plus the $u band$ PSF magnitude. Although the information in $10$ colors is in principle degenerate (i.e., could be described with just $5$ magnitudes), in order to perform a direct comparison with \cite{abraham2012}, from the $psfMag$ we derived $10$ colors. Furthermore, we also performed a few experiments on a subregion of the OPS defined by a cut in the $i$ magnitude which was forced to fall in the range $\left[ 14, 24 \right]$, plus the following cuts in colors:
\begin{itemize}
\item $-0.25 \leq u-g \leq +1.00$;
\item $-0.25 \leq g-r \leq +0.75$;
\item $ +0.30 \leq r-i \leq +0.50$;
\item $-0.30 \leq i-z \leq +0.50$.
\end{itemize}

\noindent The performance of the experiments were evaluated as percentages of the standard statistical indicators: \textit{overall efficiency}, \textit{completeness}, \textit{purity} and \textit{contamination}. With reference to the confusion matrix in Table~\ref{TAB:CM}:
\begin{table}
\centering
\begin{tabular}{c|ccc}
              &            &   OUTPUT &         \\
\hline
              &    -       & Class A  & Class B \\
TARGET        & Class A    &  $N_{AA}$ & $N_{AB}$\\
              & Class B    &  $N_{BA}$ & $N_{BB}$\\
\hline
\end{tabular}
\caption{Confusion matrix for a two class experiment. $N_{AA}$ is the number of objects in the class A correctly classified; $N_{AB}$ number of objects belonging to Class A erroneously classified in Class B; $N_{BA}$: number of objects belonging to Class B erroneously classified as belonging to Class A;  $N_{BB}$: number of objects in class B which were correctly classified.} \label{TAB:CM}
\end{table}

\begin{itemize}
\item {\it Overall Efficiency $e_{tot}$}: defined as the ratio between the number of correctly classified objects and the total number of objects in the data sets.
With reference to Table \ref{TAB:CM}

\begin{equation}
    e_{tot} \equiv \frac{N_{AA}+N_{BB}}{N_{AA}+N_{AB}+N_{BA}+N_{BB}}\label{EQ:etot}
\end{equation}

\item {\it Completeness $C_i$}: defined as the ratio between the number of correctly classified objects in a class and the total number of objects of the same class
present in the data set. Always with reference to class A:

\begin{equation}
    C_A \equiv \frac{N_{AA}}{N_{AA}+N_{AB}}\label{EQ:ca}
\end{equation}

\item {\it Purity $P_i$}: defined as the ratio between the number of correctly classified objects in a class and the total number of
objects classified in that ckass. For instance, for class A:

\begin{equation}
    P_A \equiv \frac{N_{AA}}{N_{AA}+N_{BA}}\label{EQ:pa}
\end{equation}

\item {\it Contamination $Co_i$}: defined as the dual of purity, namely as the ratio between the misclassified objects in a given class and the
number of objects classified in that class. With reference to class A:

\begin{equation}
    Co_A \equiv 1-P_A = \frac{N_{BA}}{N_{AA}+N_{BA}}\label{EQ:coa}
\end{equation}

\end{itemize}

\begin{figure}
\centering
\includegraphics[width=6cm]{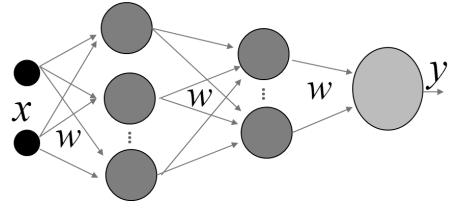}
\caption{The typical topology of a feed-forward neural network, in this case representing the architecture of MLPQNA.
In the example there are two hidden layers between the input (X) and output (Y) layers, corresponding to the architecture mostly used in the case of complex problems.}
\label{fig:mlpqna}
\end{figure}

Finally, all experiments were performed using a 2-layers MLPQNA model, i.e. a complex feed-forward architecture including two hidden layers of neurons, besides the input and output layers (Fig.~\ref{fig:mlpqna}).

\subsection{Two-class experiment: normal galaxies vs others}\label{SEC:normalgal}

\begin{table}
\centering
\begin{tabular}{lcrrr}
\hline
KB     &     step & Normal  & Others & total\\
\hline
\hline
\textit{ALL}    &  Training & 77,692  & 76,970 & 154,662 \\
       & Test      & 311,236 & 308,323&        619,559\\
\hline
\textit{CUTS}   & Training  & 77,004  & 75,915 & 152,919 \\
       & Test       & 308,512 & 304,488&        613,300\\
\hline
\end{tabular}
\caption{Characteristics of the KBs used to perform the Normal galaxies vs others experiments (from Sec.~\ref{SEC:normalgal}).} \label{TAB:twoclass_two}
\end{table}

\begin{table*}
\begin{tabular}{lccccc}
\hline
Experiment      & $\%$ $e_{tot}$ & $\%$ $C_{normal}$  & $\%$ $P_{normal}$  & $\%$ $C_{others}$  & $\%$ $P_{others}$\\
\hline
\hline
\textit{ALL}    & $89.50$   & $90.20$       & $89.03$       & $88.79$       & $89.97$\\
\textit{CUTS}   & $89.59$   & $90.44$       & $89.05$       & $88.73$       & $90.15$\\
\hline
\end{tabular}
\caption{Summary of the performance on the blind test sets for the MLPQNA experiments on the \textit{ALL} and \textit{CUTS} KBs. The values are calculated with equations from \ref{EQ:etot} to \ref{EQ:pa}.} \label{TAB:goresults}
\end{table*}

As mentioned before, a first set of experiments was performed on the $galaxy$ subset of the SDSS DR10 catalogue.
The main goal of these experiments was to evaluate whether optical photometry could be used to distinguish between normal galaxies and galaxies with a \textit{peculiar} spectrum. This is particularly important for the new surveys, which are expected to expand the amount of known AGNs, by probing them rather than QSOs only.

We therefore built a KB, based on the $10$ magnitudes ($psfMag$ and $modelMag$) available in all SDSS bands, using the objects listed in the upper side of Table \ref{TAB:KBdetails}, with also an additional flag, set to $0$ for the normal galaxies and $1$ for all other types.

With reference to Table \ref{TAB:KBdetails} we see that in the whole KB there are $1,911,513$ objects among which $1,526,215$ normal galaxies and $385,298$ objects belonging to the other classes.
In order to equally weight the two groups, we randomly extracted $\sim25\%$ of the galaxies obtaining a smaller data set. We also cleaned the KB by removing Not-a-Number (NaN) entries.
The resulting data set (hereafter named \textit{ALL}) contained $619,559$ objects among which $311,236$ were flagged as normal galaxies. Furthermore, in order to exclude objects falling in the poorly populated parts of the OPS, we also created a second data set by applying some cuts in magnitudes, i.e. by removing the very low significant tails of the magnitude distributions, consisting of $\sim1\%$ of the original data.
This second KB will be referred to as \textit{CUTS} in what follows. The properties of the two different KBs and the way they were split into training and test set, are listed in Table~\ref{TAB:twoclass_two}.

The main results of the classification with the MLPQNA model are summarised in Table~\ref{TAB:goresults}, while in Tables \ref{TAB:misALL} and \ref{TAB:misCUTS} are listed, respectively, the misclassified objects for the \textit{ALL} and for the \textit{CUTS} experiments, divided according to the SDSS spectroscopic subclasses.

The first thing to be noted from the results of Table~\ref{TAB:goresults} is that the performance of the MLPQNA on the two different data sets differ only by a negligible amount, although with a slightly better behavior in the \textit{CUTS}. Nonetheless we considered the \textit{ALL} experiment as the best one, because it does not require any restriction on the OPS. This induced us to exclude any cut in further experiments.

Looking at Tables \ref{TAB:misALL} and \ref{TAB:misCUTS}, which summarize the results of these experiments, one notices that all objects labeled as \textit{broadline} are poorly classified (with contamination ranging from $69\%$ to $30\%$). This fact induced us to check whether a different partition of the spectroscopic subclasses could lead to some improvements in the performance.

We therefore performed two additional experiments on the \textit{ALL} data set. In the first case by grouping the objects in two broad classes: NORMAL galaxies plus BROADLINE types vs others (hereafter \textit{NBonly} experiment); in the second case by grouping together all the BROADLINE types vs others (hereafter \textit{NBall} experiment).

Finally, always in order to minimize the number of misclassified objects, an additional $two-class$ classification experiment was performed by splitting the data set in two classes: one containing NORMAL, BROADLINE, AGN and AGN BROADLINE, and the other including all remaining types (hereafter experiment \textit{NBA}).

The results for the three experiments are summarized in Table~\ref{TAB:NB}, while the percentage of misclassified objects in the three cases are reported, respectively, in the Table~\ref{TAB:NBonly} for \textit{NBonly}, Table~\ref{TAB:NBall} for \textit{NBall} and Table~\ref{TAB:NBA} for \textit{NBA}.

\begin{table*}
\begin{tabular}{lccccc}
\hline
Class      & Subclass         & N. objects & N. objects & misclassified\\
           &                  &Subclass    & class      & $\%$\\
\hline
\hline
NORMAL     &                  &         & 311,236&10\\
\hline
           & AGN              & 18,119  &        & 41\\
           & BROADLINE        & 14,936  &        & 67\\
		   & STARBURST        & 58,371  &        & 2 \\
OTHERS 	   & STARFORMING      &212,748  &308,323 & 7\\
           & AGN BROADLINE    &2,798    &        & 47\\
           & STARBURST BROAD  &133      &        & 14\\
		   & STARFORMING BROAD&1,218    &        & 30\\
\hline
\end{tabular}
\caption{Summary of misclassified objects in the \textit{ALL} experiment. Column $1$: SDSS spectroscopic class; column $2$: SDSS spectroscopic subclass; column $3$ and $4$: number of objects in the subclass and in the class; column $5$: fraction of misclassified objects.} \label{TAB:misALL}
\end{table*}

\begin{table*}
\begin{tabular}{lccccc}
\hline
Class      & Subclass         & N. objects & N. objects & misclassified\\
           &                  &Subclass    & class      & $\%$\\
\hline
\hline
NORMAL     &                  &         & 308,512&10\\
\hline
           & AGN              & 17,971  &        & 41\\
 		   & BROADLINE        & 14,356  &        & 69\\
		   & STARBURST        & 56,922  &        & 2 \\
OTHERS	   & STARFORMING      &211,238  &304,488 & 7\\
           & AGN BROADLINE    &2,688    &        & 48\\
           & STARBURST BROAD  &132      &        & 15\\
		   & STARFORMING BROAD&1,181    &        & 31\\
\hline
\end{tabular}
\caption{Summary of misclassified objects in the \textit{CUTS} experiment. Columns as in table \ref{TAB:misALL}.} \label{TAB:misCUTS}
\end{table*}

\begin{table*}
\begin{tabular}{lccccc}
\hline
Class      & Subclass         & N. objects & N. objects & misclassified\\
           &                  &Subclass    & class      & $\%$\\
\hline
\hline
NORMAL                          & NORMAL            & 278,213 & 292,850                    &  9\\
+ BROADLINE                     & BROADLINE         & 14,637  &                            &  9\\
\hline
\multirow{6}{*}{OTHERS}         &AGN                & 17,750  &\multirow{6}{*}{285,104}    & 46\\
                                & STARBURST         & 56,443  &                            &  2\\
					            & STARFORMING       & 206,853 &                            &  7\\
                                & AGN BROADLINE     & 2,715   &                            & 67\\
					            & STARBURST BROAD   & 133     &                            & 17\\
                                & STARFORMING BROAD & 1,210   &                            & 41\\
\hline
\end{tabular}
\caption{Summary of misclassified objects in the \textit{NBonly} experiment. Columns as in table \ref{TAB:misALL}.} \label{TAB:NBonly}
\end{table*}

\begin{table*}
\begin{tabular}{lccccc}
\hline
Class      & Subclass         & N. objects & N. objects & misclassified\\
           &                  &Subclass    & class      & $\%$\\
\hline
\hline
                & NORMAL            & 277,977   &            & 9\\
NORMAL          & BROADLINE         & 14,626    &            & 7\\
+ \textit{ALL}  & AGN BROADLINE     & 2,764     &  296,691   &31\\
BROADLINE		& STARBURST BROAD   & 127       &            &84 \\
                & STARFORMING BROAD & 1,197     &            &51\\
\hline
                &AGN                & 17,633    &            & 48\\
OTHERS          & STARBURST         & 56,484    &   280,657  &  2\\
				& STARFORMING       & 206,540   &            &  7\\
\hline
\end{tabular}
\caption{Summary of misclassified objects in the \textit{NBall} experiment. Columns as in table \ref{TAB:misALL}.} \label{TAB:NBall}
\end{table*}

\begin{table*}
\begin{tabular}{lccccc}
\hline
Class      & Subclass         & N. objects & N. objects & misclassified\\
           &                  &Subclass    & class      & $\%$\\
\hline
\hline
NORMAL                       & NORMAL               & 236,847 & \multirow{4}{*}{272,700}     & 8\\
+ BROADLINE                  & BROADLINE            & 14,356  &                              & 5\\
+ AGN                        & AGN                  & 18,119  &                              & 43 \\
+ AGN BROAD                  & AGN BROADLINE        &  2,798  &                              & 23\\
\hline
\multirow{4}{*}{OTHERS}      & STARBURST            & 58,371  &    \multirow{4}{*}{272,470}  & 2\\
                             & STARFORMING          & 212,748 &                              & 9\\
					         & STARBURST BROAD      &   133   &                              & 9\\
                             & STARFORMING BROAD    &  1,218  &                              & 54\\

\hline
\end{tabular}
\caption{Summary of misclassified objects in the \textit{NBA} experiment. Columns as in table \ref{TAB:misALL}.} \label{TAB:NBA}
\end{table*}

\begin{table*}
\begin{tabular}{ccccc}
\hline
Experiment          & $\%$ $e_{tot}$        & class     & \% completeness     & \% purity \\
\hline
\hline
\multirow{2}{*}{\textit{NBonly}}   &   \multirow{2}{*}{91.03}        & NORMAL+BROADLINE  & 91.25  & 91.06\\
\cline{3-5}
& & OTHERS &      90.80   & 90.99\\
\hline
\hline
\multirow{2}{*}{\textit{NBall}} &   \multirow{2}{*}{91.11}          & NORMAL+\textit{ALL} BROADLINE  & 91.07  & 91.57 \\
\cline{3-5}
& & OTHERS     & 91.14    & 90.62\\
\hline
\hline
\multirow{2}{*}{\textit{NBA}} &  \multirow{2}{*}{91.17}            & NORMAL+BROADLINE+\textit{ALL} AGN     & 89.88     & 92.27\\
\cline{3-5}
& & OTHERS & 92.46  & 90.13\\
\hline
\hline
\end{tabular}
\caption{Summary of the statistical performance on the two-class experiments named as \textit{NBonly}, \textit{NBall} and \textit{NBA}. The columns $2$ and $5$ are calculated with equations from \ref{EQ:etot} to \ref{EQ:pa}} \label{TAB:NB}
\end{table*}

\subsection{Two-class experiment: QSO vs Star}\label{SEC:twoclass}

The identification of quasar candidates on photometric grounds is a topic of the highest relevance. In order to evaluate the best possible combination of input features and model architecture, we run several experiments but, given the large computational load, it was not advisable to run the experiments on the whole KB and therefore, for these exploratory experiments, we used a reduced KB randomly extracted from the original one.
Furthermore, in terms of taxonomy of all the combinations among magnitudes and colors, we were also interested in evaluating performance in comparison with the experiment types described in \cite{abraham2012}. Hence, only colors have been used. The performed experiments were indeed composed by the following cases:

\begin{description}
  \item[a)] two experiments with $4$ colors and one magnitude, respectively, $u$ and $r$ (hereinafter named as, respectively, 2a-u and 2a-r);
  \item[b)] as with $2a$, but with all same cuts in magnitudes and colors as in \cite{abraham2012}, (hereinafter named as, respectively, 2b-u and 2b-r);
  \item[c)] as with $2a$, but with only a cut in the $i$ magnitude, by taking into account what was done by \cite{abraham2012}, (hereinafter named as, respectively, 2c-u and 2c-r);
  \item[d)] as with $2a$, but using $10$ colors, obtained by combinations of $psfMag$ magnitudes, (hereinafter named as, respectively, 2d-u and 2d-r);
  \item[e)] as with $2b$, extended to $10$ colors, (hereinafter named as, respectively, 2e-u and 2e-r);
  \item[f)] as with $2c$, extended to $10$ colors, (hereinafter named as, respectively, 2f-u and 2f-r).
\end{description}

\begin{table*}
\centering
\begin{tabular}{lcccccc}
\hline
EXP ID  &   $\%$ $e_{tot}$  &   Star $\% C_S$    &   Star $\% P_S$ &   QSO $\% C_Q$    &   QSO $\% P_Q$ \\
\hline
\hline
2a-u    &   $91.24$         &   $87.76$          &   $95.23$       &   $95.11$         &   $87.48$ \\
2a-r    &   $91.30$         &   $87.92$          &   $95.18$       &   $95.05$         &   $87.63$ \\
2b-u    &   $88.06$         &   $78.71$          &   $98.72$       &   $98.04$         &   $81.19$ \\
2b-r    &   $88.13$         &   $79.73$          &   $98.92$       &   $98.21$         &   $81.96$ \\
2c-u    &   $91.22$         &   $87.20$          &   $95.68$       &   $95.56$         &   $87.13$ \\
2c-r    &   $91.15$         &   $86.25$          &   $96.47$       &   $96.53$         &   $86.47$ \\
2d-u    &   $91.34$         &   $88.09$          &   $95.06$       &   $94.93$         &   $87.81$ \\
2d-r*   &   $91.42$         &   $88.08$          &   $95.23$       &   $95.11$         &   $87.82$ \\
2e-u    &   $89.15$         &   $80.52$          &   $98.39$       &   $98.56$         &   $82.28$ \\
2e-r    &   $89.05$         &   $80.10$          &   $98.61$       &   $98.77$         &   $82.05$ \\
2f-u    &   $91.42$         &   $88.49$          &   $95.23$       &   $94.83$         &   $87.61$ \\
2f-r    &   $91.43$         &   $88.67$          &   $95.16$       &   $94.68$         &   $87.63$ \\
 \hline
\end{tabular}
\caption{Summary of the results of the two-class experiments (from Sec.~\ref{SEC:twoclass}), where the best is indicated with the asterisk (referenced in the text as 2d-r) referred to the parameter space composed by the $10$ colors without any cut for each object. The training and test sets are respectively $10\%$ and $90\%$ of the given dataset. The last five columns are referred to equations from \ref{EQ:etot} to \ref{EQ:pa}. All the quantities reported in the table are percentages.}
\label{TAB:twoclass}
\end{table*}

The results of the experiment are reported in Table~\ref{TAB:twoclass}.
In statistical terms, the experiments with the $i$ cut resulted comparable to those without any cut. Hence, we considered the latter group as the best.
These results will be further discussed in Sect. \ref{SEC:conclusions}.

\subsection{Three-class experiments}\label{SEC:threeclass}
This set of experiments aimed at reproducing on photometric grounds the SDSS spectroscopic classification in the three main classes (STAR, GALAXY, QSO). We therefore performed our $three-class$ experiments using:

\begin{description}
  \item[a)] $10$ magnitudes (hereinafter named as 3a), composed by the five $magModel$ plus the five $psfMag$, without any photometric cut, in order to properly weight the contribution from the nuclear regions;
  \item[b)] $10$ magnitudes composed by the five $magModel$ and the five $psfMag$ with a cut in the $i$ magnitude (hereinafter named as 3b);
  \item[c)] $10$ colors from $psfMag$ type and a $magModel$ magnitude, respectively, $u$ and $r$ (hereinafter named as, respectively, 3c-u and 3c-r);
  \item[d)] $5$ magnitudes, alternately from $psfMag$ and $magModel$ (hereinafter named respectively as 3d-psf and 3d-model), in order to evaluate the single contribution of the two types.
\end{description}

In all cases the training set was randomly extracted by balancing the number of examples presented to the network for each class, while the given dataset has been always randomly split into a training and a blind test sets, by using percentages of respectively, $12\%$ and $88\%$.

\begin{table*}
\centering
\begin{tabular}{lcccccccc}
\hline
EXP ID      &   $\% e_{tot}$    &   Galaxy $\% C_G$     &   Galaxy $\% P_G$ &   QSO $\% C_Q$    &   QSO $\% P_Q$    &   Star $\% C_S$   &   Star $\% P_S$ \\
\hline
\hline
3a*         &   $91.31$         &   $97.02$             &   $93.49$         &   $90.49$         &   $86.90$         &   $86.40$         &   $93.82$\\
3b          &   $91.02$         &   $96.96$             &   $93.49$         &   $88.95$         &   $86.84$         &   $87.19$         &   $92.89$\\
3c-u        &   $87.83$         &   $92.69$             &   $88.00$         &   $88.27$         &   $85.56$         &   $82.57$         &   $90.21$\\
3c-r        &   $87.77$         &   $92.64$             &   $88.03$         &   $88.42$         &   $85.60$         &   $82.27$         &   $89.93$\\
3d-psf      &   $86.62$         &   $90.73$             &   $86.58$         &   $87.94$         &   $83.28$         &   $81.23$         &   $90.65$\\
3d-model    &   $87.64$         &   $93.60$             &   $87.32$         &   $88.13$         &   $85.76$         &   $81.23$         &   $90.17$\\
\hline
\end{tabular}
\caption{Summary of the results of the three-class experiments (from Sec.~\ref{SEC:threeclass}). The best one is with the asterisk (referenced in the text as 3a), referred to the parameter space composed by the $10$ magnitudes ($psfMag$ and $magModel$) for each object. The training and test sets are respectively $12\%$ and $88\%$ of the given dataset. The columns are referred to equations from \ref{EQ:etot} to \ref{EQ:pa}. All the quantities reported in the table are percentages.}
\label{TAB:threeclass}
\end{table*}

The results of the $three-class$ experiments are reported in Table~\ref{TAB:threeclass}. By looking at the QSO statistics, the purity/completeness results of $3a$ experiment appeared very similar to those obtained in the $3b$ experiment. Although these two experiments led to comparable results, we consider the $3a$ as the best, because obtained without any cut in the parameters. For a complete discussion see Sect. \ref{SEC:conclusions_1}.

\section{Discussion}\label{SEC:conclusions}

The classification experiments described in the previous sections aimed at producing a reliable catalogue of candidate QSOs/AGNs by exploiting the photometric information available within the DR10. In what follows we give a detailed overview of their results.

\subsection{Galaxy versus others}\label{SEC:galvsothe}
The set of experiments described in Sec.~\ref{SEC:normalgal} aimed at investigating whether optical photometry could be used to isolate different types of galaxies as defined by the spectroscopic subclasses in the SDSS DR10 spectroscopic archive.
With reference to Table~\ref{TAB:KBdetails} it is apparent that different types of objects are represented in the KB in a very uneven way. In the first couple of experiments (\textit{ALL} and \textit{CUTS}) we tried to separate normal galaxies from those having a \textit{peculiar} spectrum (i.e. all the other subclasses in the SDSS classification scheme). The performance of the method, shown in Table~\ref{TAB:goresults}, were not affected by the application of color and magnitude cuts and the efficiency remained stationary around a value of $\sim90\%$. Given the lack of improvements between the \textit{ALL} and \textit{CUTS} experiments, we decided to perform supplementary experiments by freezing the \textit{ALL} KB, the same configuration for the network and the input parameters.

The statistics of misclassified objects, reported in Tables \ref{TAB:misALL} and \ref{TAB:misCUTS}, show that the MLPQNA is quite effective in disentangling starburst and starforming galaxies both narrow and broadline (misclassification rates of $2\%$ and $7\%$, respectively), and much less effective in identifying other types of objects such as, for instance, AGN (misclassification rate $41\%$) and AGN broadline objects ($\sim47\%$). The performance on the starburst broadline and starforming broadline types, however, are difficult to evaluate, since these two groups are very poorly represented in the KB and therefore the network may have failed in capture their specific signatures in the OPS.

The fact that $41\%$ of the AGN, $67\%$ of the broadline and $47\%$ of the AGN broadline were misclassified induced us to check whether different groupings of the subclasses could lead to improvements in the classification efficiency.

In the \textit{NBonly} experiment (Table~\ref{TAB:NBonly}) the misclassification rate improved to $9\%$ for the normal galaxies and dropped to $9\%$ (from $67\%$) for the broadline. The misclassification rates for AGN broadline, starburst broadline and starforming broadline increased, confirming the fact that the network is not capable to disentangle these spectroscopic subclasses in an effective way.
In the \textit{NBall} experiment (Table~\ref{TAB:NBall}) the results did not change in a significant way. In fact, the normal, starburst and starforming type objects still maintain a misclassification rate of, respectively, $9\%$, $2\%$ and $7\%$. While for the broadline galaxies the misclassification decreases to $7\%$. A slightly better result is obtained for the AGN broadline type, going from $47\%$ in \textit{ALL} and $67\%$ in \textit{NBonly}, to $31\%$. The AGN and starforming broadline types are even worse reaching a misclassification of, respectively $48\%$ and $51\%$. Finally, the starburst broadline, although sparsely represented in the data set, resulted almost completely misclassified ($84\%$).

The last experiment (\textit{NBA}), resulted as the best one in terms of overall efficiency as well as for classes normal, broadline, AGN broadline and starburst broadline, where the misclassification rates decreased to, respectively, $8\%$, $5\%$, $23\%$ and $9\%$. Only starburst type rate remains unchanged ($2\%$), while AGN objects perform slightly worse ($43\%$ vs $41\%$ obtained in the \textit{ALL} case). On the other hand, the classification of starforming and starforming broadline type decreases to, respectively, $9\%$ and $54\%$, although the starforming rate decreases by $2\%$ only. The strong decrease for the starforming broadline type could be due to its poor presence within the data set.

\subsection{QSO versus Stars}
Already in the earliest SDSS works \citep{stoughton2002,richards2002} optical colors were used to disentangle stars from quasars in the SDSS color space. \cite{dabrusco2009} used an unsupervised clustering algorithm followed by an agglomeration phase to identify candidate QSOs in a parameter space based only on photometric colors.

\citep{sinha2007,abraham2012} used a Difference Boosting Neural Network (DBNN) on SDSS optical colors only, achieving excellent performance ($\sim98\%$) in disentangling QSO from normal galaxies in the unresolved source catalogues (thus including stars, quasars and unresolved galaxies). They however applied a cut in colors by isolating the region of this space where most ($84\%$) of the spectroscopically confirmed quasars lay. This choice, however, penalizes the recognition of peculiar objects which are likely to lay outside of the main distribution of normal \textit{well behaved} objects.

Our experiments are described in Sec.~\ref{SEC:twoclass} and related results are shown in Table~\ref{TAB:twoclass}. The immediate conclusion which can be drawn by analyzing the results is that our method is able to achieve $\sim95\%$ of completeness and a purity of $\sim88\%$ without having to apply any color cuts.

\subsection{Star - Galaxy - QSO}\label{SEC:conclusions_1}

In the three-class classification experiment the KB included all objects in the spectroscopic catalogue regardless their extended or unresolved nature.

The first thing to notice is that all experiments reported in Table~\ref{TAB:threeclass} have performance which differ only by very little amount. The experiment which led to the best results was the $3a$ with an overall efficiency of $\sim91\%$.
This experiment was performed using a $2$ layers MLPQNA and a quite small training set (only $12\%$ of the available KB) using as input parameters the $10$ SDSS magnitudes.
Having identified the experiment $3a$ as the best one, two additional experiments were performed on the $3a$ parameter space, by changing either the topology of the neural network or the amount of training set to evaluate their individual contribution to the performance. In the first case we used a $1$ layer MLPQNA and in a second case a larger training set ($60\%$ of the KB). The case with the increased training set led to a negligible improvement in the overall efficiency but at the price of a huge increase in computing time. While, in the case of a single layer MLPQNA network, the performance decreased of about $1\%$.
This can be understood by taking into account that the complexity introduced by the second hidden layer in the architecture of the neural network is always justified when the KB samples the OPS only sparsely. In the case discussed here, even with a reduced KB, the number of samples in the training set was sufficiently high to ensure a proper coverage of the OPS and to minimize the requirements on generalization capabilities.
The additional complexity introduced by the second layer proves indeed very useful in presence of much smaller KBs \citep{cavuoti2012}. Surprisingly enough, the introduction of mag $i$ cuts in the parameter space (experiment $3b$) did not lead to any significant improvement in the method efficiency.

In summary, using optical data only, the MLPQNA seems quite effective in identifying galaxies (completeness $97.02\%$ and contamination $6.51\%$) from QSO and stars (Table~\ref{TAB:threeclass}).

We performed the supervised learning classification experiment to produce a catalogue of high fidelity candidate QSO selected on photometric data only. The classification is based on a membership probability criterium: for each object the model output is based on three different membership pseudo-probabilities (a confidence level for each class). Therefore the values in Table~\ref{TAB:threeclass} are referred to these confidence levels for all objects of the blind test set (technique also known as \textit{winner-takes-all}; \citealt{grossberg1973}). With this method, in the case of the selected best experiment, we reached a QSO class purity of $\sim 87\%$. Afterwards, since our goal was to reach the highest level of purity in the produced catalogue, we performed a further statistical analysis of the test set starting from the results of the experiment $3a$, by assessing the variation of purity vs completeness as a function of the increasing confidence threshold used to evaluate the QSO candidates from the trained MLPQNA model output. At the end we reached a purity of $\sim95\%$ in the blind test set, at the price of a reduced completeness. The resulting QSO photometric catalogue contains $3,602,210$ candidates and will be made publicly available through the CDS VizieR facility.

We wish to emphasize that a three-class classification approach offers a significant advantage with respect to the traditional approach based on serialized series of two-class classification steps, since it does not introduce systematic biases due to objects which are misclassified at each step of the sequence.
Specifically: the separation among star, galaxies and quasars could be achieved via two independent classification steps: first separating resolved (galaxies) from unresolved objects (stars and quasars) and then dividing the latter group into stars and quasars. Objects misclassified at the first step would be propagated into the next step.

Restricting ourselves to the QSO vs STAR classification, in the three-class experiment we achieve an overall accuracy of $91.5\%$, with a QSO purity of  $88.4\%$ and QSO completeness of $95.3\%$. Therefore, by comparing this result with those of the two class experiment (Table~\ref{TAB:twoclass}) there is a noticeable improvement of accuracy in the three-class experiment. This confirms the fact that a three-class approach is preferable to the hierarchical chain of two-class experiments.  The slight decrease in QSO completeness found in the three-class experiment may be simply due to the higher separation complexity of the experiment.

It needs to be stressed, however, that the catalogue of candidate quasars needs to be used with some cautions. In fact, no a-priori defined photometric cuts have been applied to the data and our operative definition of a quasar was induced only by the properties of the spectroscopic knowledge base. Hence, any bias, potentially present in the KB, would be reflected in the final definition of the catalogue.

Let us, for instance, take into account the bright end of the luminosity distribution. It is well known that the SDSS quasar catalogue is fairly ($\sim 95 \%$) complete for $i < 19.1$ \citep{winchatz2007,{ross2012}}.
If we consider the produced catalogue, within the flux limit of $i<19.1$ we find $\sim50\%$ more candidates than those present in the spectroscopic sample. But, as already discussed in the literature \citep{croom2009}, this bright end is strongly contaminated by stars, UVX sources and, mainly, by narrow emission-line galaxies. In order to isolate luminous quasars, \cite{croom2009} used a complex system of photometric cuts:

\begin{description}
  \item[A)] u-g$<0.8 \bigcap$ g-r$<0.6 \bigcap$ r-i$<0.6$;
  \item[B)] u-g$>0.6 \bigcap$ g-i$>0.2$;
  \item[C)] u-g$>0.45 \bigcap$ g-i$>0.35$;
  \item[D)] galprob$>0.99 \bigcap$ u-g$>0.2 \bigcap$ g-r$>0.25 \bigcap$ r-i$<0.3$;
  \item[E)] galprob$>0.99 \bigcap$ u-g$>0.45$.
\end{description}

These cuts were indeed combined through the logical expression [$A \bigcap \overline{B} \bigcap \overline{C} \bigcap \overline{D} \bigcap \overline{E}$], to carve the optimal quasar locus in the observed parameter space.

If we apply the above conditions to both spectroscopic KB and the final catalogue of candidate quasars, under the flux limit of $i<19.1$ the overabundance of objects in the produced catalogue is reduced of the $\sim17\%$.

In other words, while ML methods prove very effective in partitioning the OPS accordingly to the information contained in the spectroscopic KB, the correct interpretation of their output needs to be fine tuned using the expert's (i.e. the astronomer's) knowledge. For this reason, in order to allow interested readers to apply their own filters to the data, the catalogue contains all the relevant photometric information.

\begin{figure*}
\centering
\includegraphics[width=16cm]{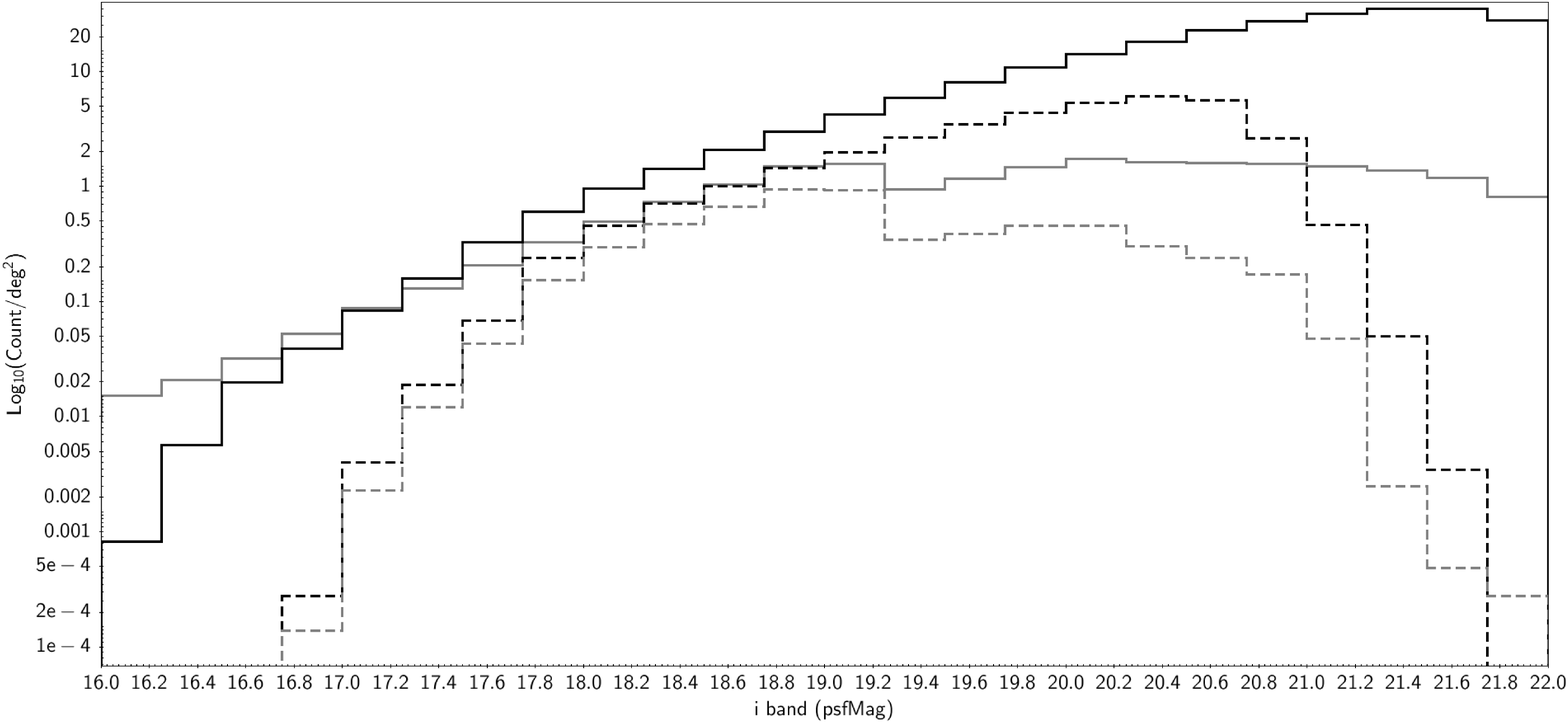}
\caption{Logarithmic distribution of the number of objects per square degree as a function of $psfMag\_i$ magnitude, before (solid line) and after (dashed line) the quality cut specified by Croom et al. (2009). Black lines refer to our catalogue, gray lines to the SDSS spectroscopic sample. The solid black line represents the resulting catalogue of candidate quasars/AGNs, consisting of $\sim 3.6$ million objects. While the dashed black line indicates the $\sim0.5$ million objects flagged as robust candidates.}
\label{fig:histo1}
\end{figure*}

In Fig.~\ref{fig:histo1} we compare the number of candidate QSOs (solid black line) in our catalogue vs the number of objects in the KB which are spectroscopically identified as QSOs (solid grey line), both before and after (dashed lines) the cuts of \cite{croom2009}. Before the cuts, the excess of spectroscopically confirmed AGNs/QSOs at bright magnitudes ($i<17$) can be easily explained by the fact that low luminosity AGNs, which are spectroscopically identified, can easily escape photometric detection since the relative weight of the AGN contribution is negligible with respect to the contribution of the central regions of the galaxy. Furthermore, in this magnitude range the SDSS sample is highly unbalanced, thus causing the MLPQNA model to be less accurate in terms of training performance on the QSO class.
After applying the \cite{croom2009} selection criteria, however, the agreement between the two distributions becomes remarkably good and it is maintained down to the completeness limit of the SDSS spectroscopic sample ($i = 19.1$).
Beyond such limit, the two distributions begin to differ due to both new candidate quasars which were not spectroscopically confirmed in the SDSS (region between the two dashed lines) and to a contamination from low/medium luminosity AGNs (region between the two black lines) which, as it has been discussed in Sec.~\ref{SEC:galvsothe}, cannot be disentangled on the grounds of optical photometry only.

\begin{figure*}
\centering
\includegraphics[width=8cm]{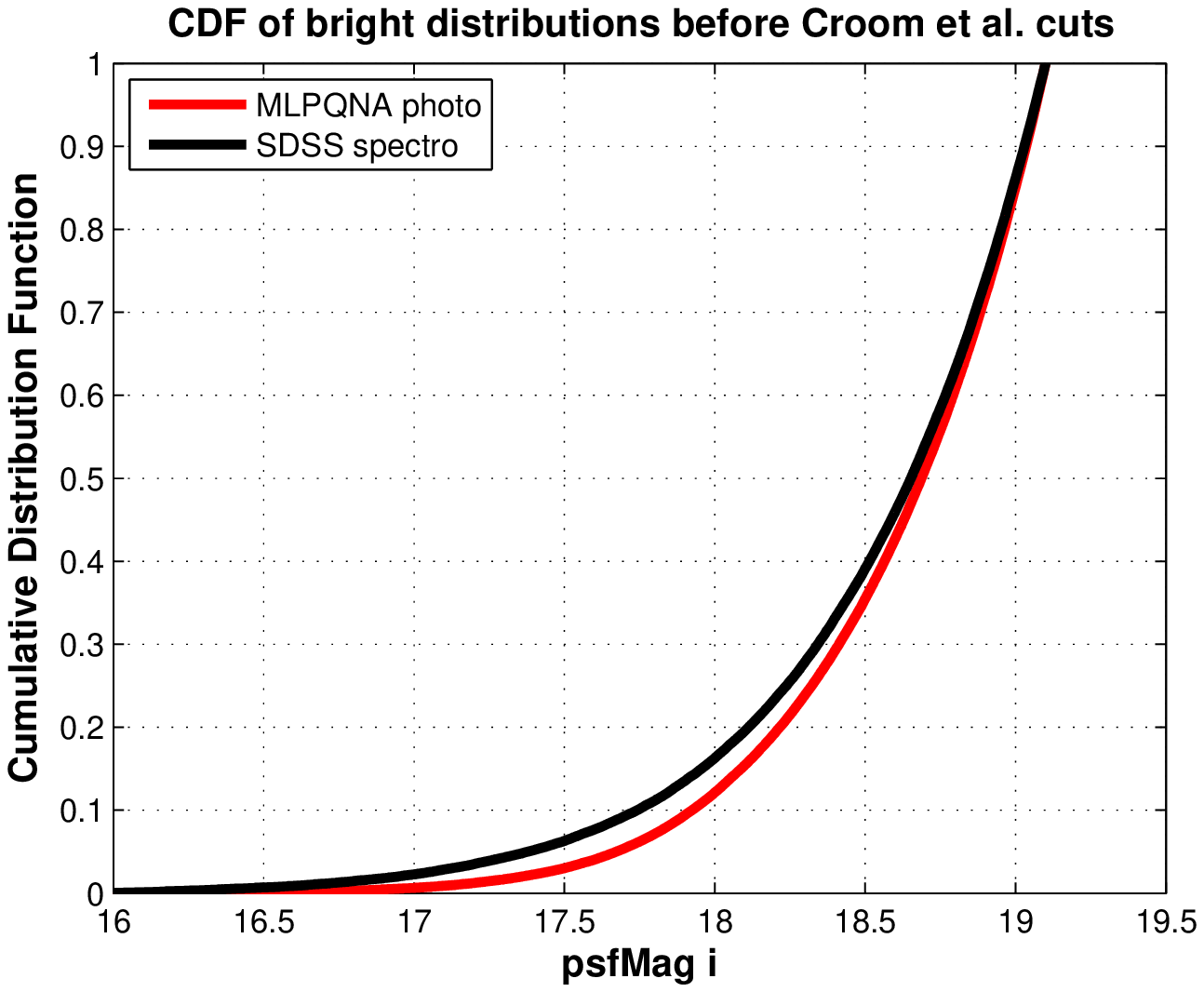}
\includegraphics[width=8cm]{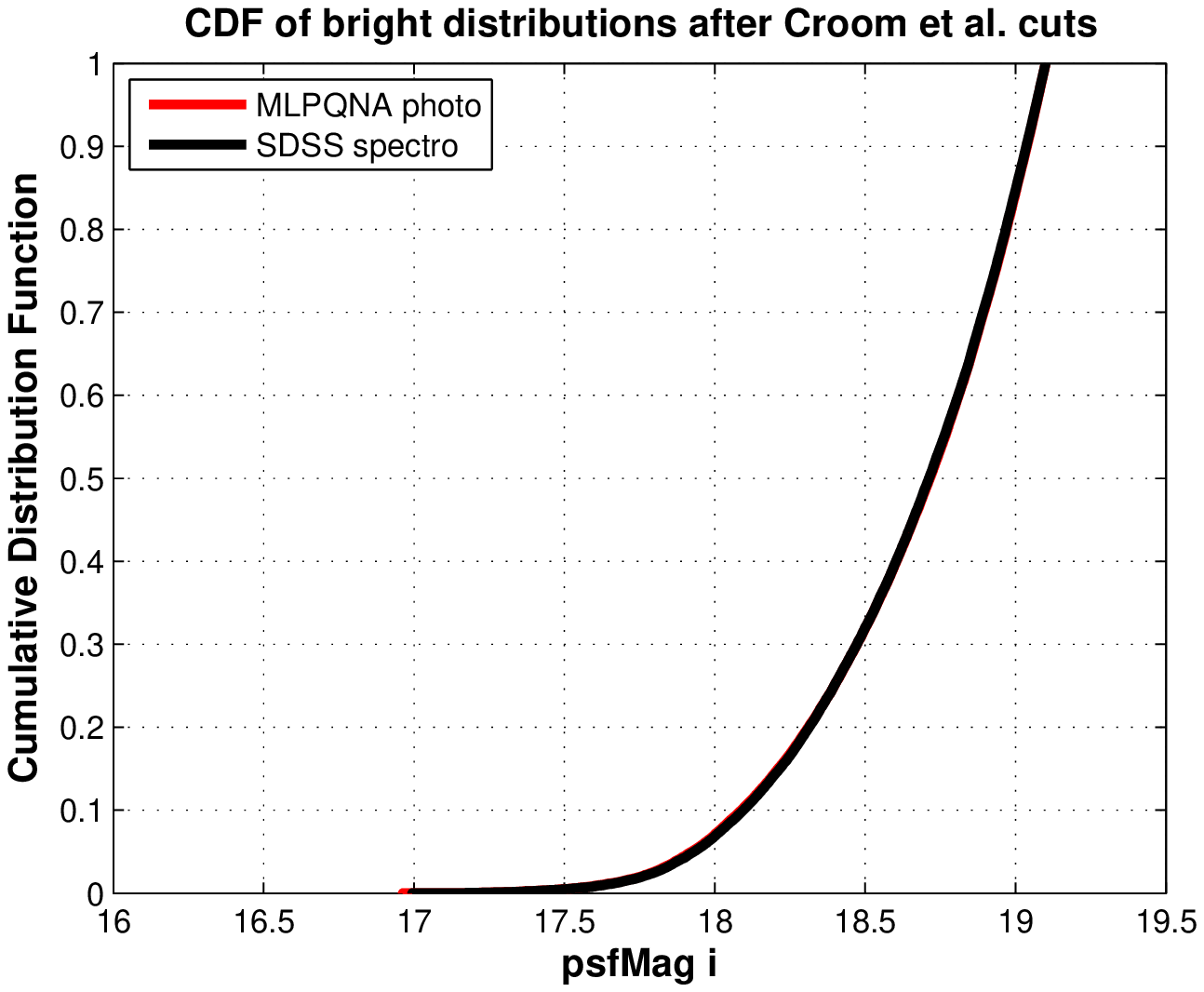}
\includegraphics[width=8cm]{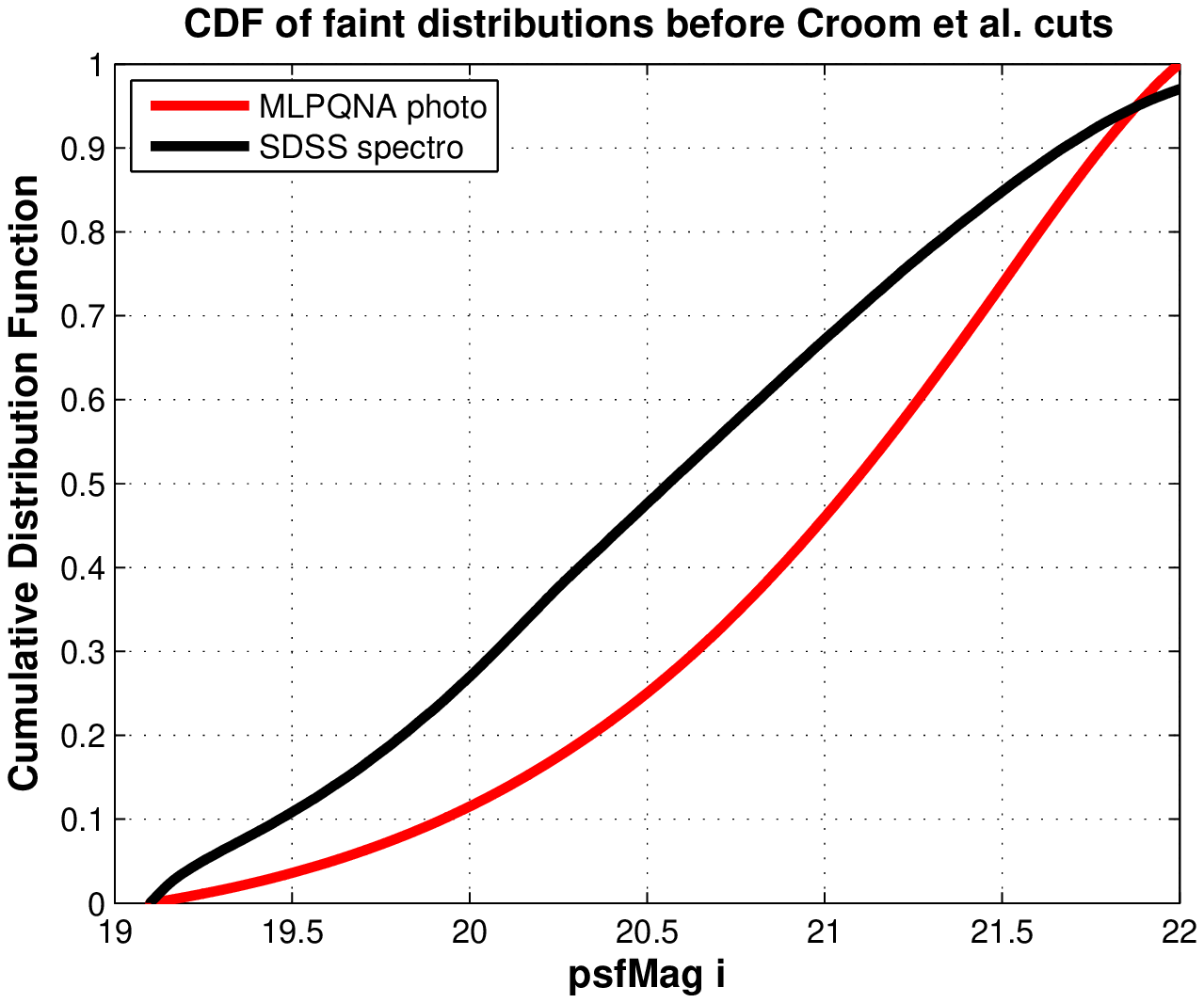}
\includegraphics[width=8cm]{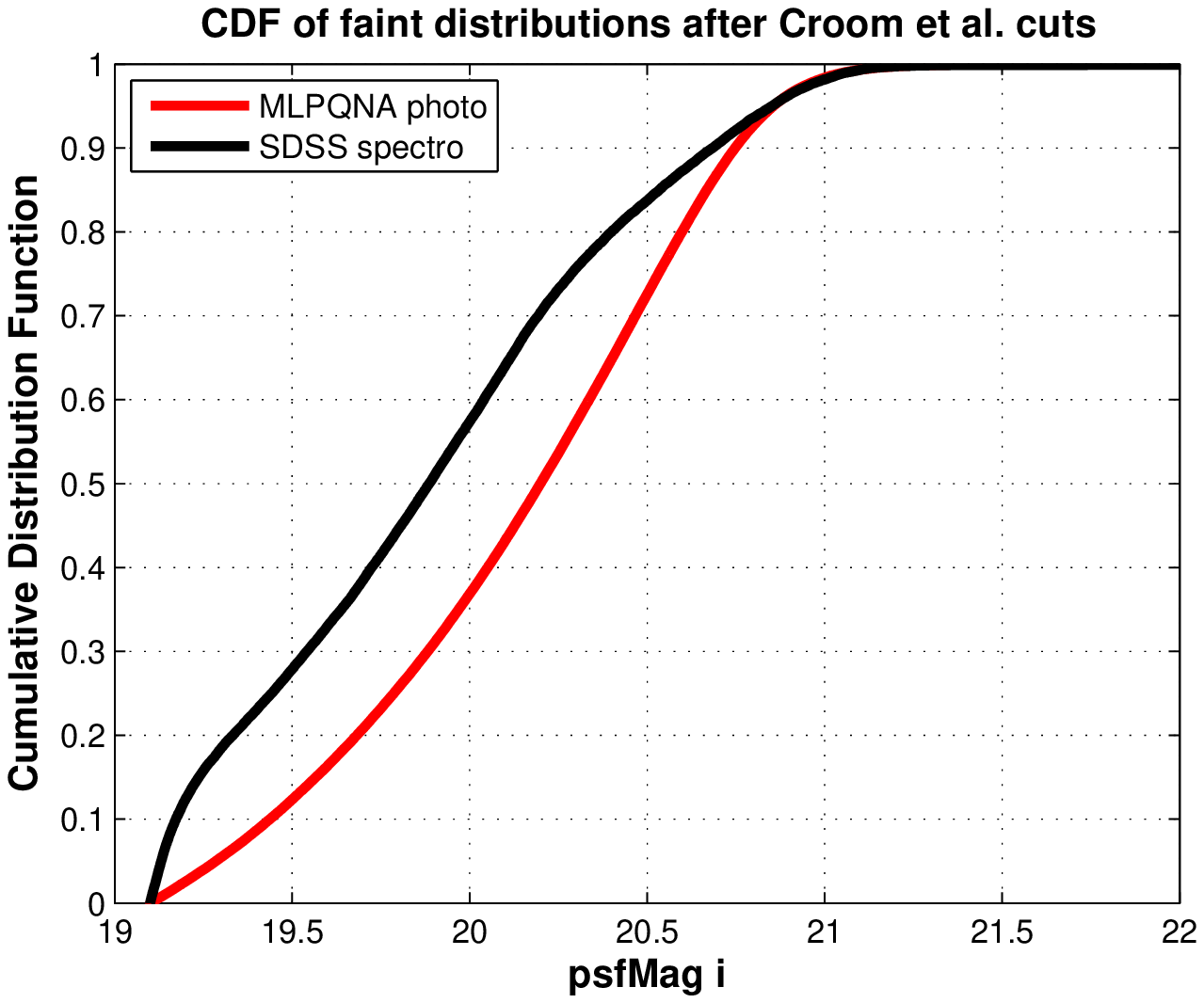}
\caption{Results of the 1D Kolmogorov-Smirnov test to compare the $psfMag\_i$ band distributions of QSO/AGN candidates between the photometric catalogue (red line) and the spectroscopic KB (black line). The upper side diagrams show the tests done for brighter sources ($i < 19.1$), respectively, before (upper left) and after (upper right) the cuts of Croom et al. (2009). The lower side reports the same type diagrams for fainter sources ($i \geq 19.1$). The K-S test resulted positive in the two cases shown in the upper side diagrams, confirming the similarity at the $95\%$ of significance level within the spectroscopic completeness limit. In particular, the two distributions appear perfectly overlapped in the upper right diagram.}
\label{fig:CDF4}
\end{figure*}

The trends depicted in the diagram of Fig.~\ref{fig:histo1} find a statistical confirmation in Fig.~\ref{fig:CDF4}, where we show the results of the 1D Kolmogorov-Smirnov (K-S) test \citep{ledermann1982}, graphically represented through the cumulative $psfMag\_{i}$ band distribution functions. The test has been performed to compare the distributions of QSO/AGN candidate sources, respectively, between the photometric catalog and the spectroscopic KB, before and after the completeness limit of the SDSS spectroscopic sample ($i = 19.1$) and the \cite{croom2009} cuts application. Within the completeness limit, the K-S test confirms the similarity of the two distributions, perfectly overlapping after the application of \cite{croom2009} cuts. Beyond the completeness limit, the K-S test makes evident a difference between the two distributions, considered reasonable by taking into account the incompleteness of the spectroscopic sample.

In the produced catalogue, a quality flag has been included to take the \cite{croom2009} cuts into consideration.

\section{Conclusions}
The main scope of the present work was to investigate the possibility of disentangling different spectral classes of objects, by exploiting the machine learning based model named MLPQNA \citep{brescia2012}. In order to reach these goals, three categories of experiments have been performed.

In a first group of experiments we investigated the possibility to disentangle different spectroscopic types present in the GALAXY class. All experiments, summarized in the Tables \ref{TAB:goresults} and \ref{TAB:NB} were unable to separate AGNs from normal galaxies on the grounds of optical data only, although the overall efficiency is always $\sim90\%$, thus confirming what we found in \cite{cavuoti2014}. By excluding the AGN class, in all other cases where the number of representative objects in the KB is sufficiently large, the disentangling appears quite feasible, in particular by considering results of the experiment \textit{NBA} (Table~\ref{TAB:NBA}).

The second one concerned distinguishing between QSO and STAR classes. The results, shown in Table~\ref{TAB:twoclass}, confirmed the findings of \cite{abraham2012}, that the use of $10$ colors, rather than the standard $4$ colors, although without carrying a great amount of additional information, may help machine learning methods to find better solutions.

The third category focused on identifying candidate QSOs from the whole catalogue including also stars and galaxies, thus permitting to release the candidate QSO photometric catalogue. In this case, the possibility of avoiding the known downside of hierarchical pairwise classification (multiplicative propagation of misclassification), induced an intrinsic advantage of a direct classification of QSOs from the whole catalogue. The resulting catalogue of QSO candidates contains $3,602,210$ objects, of which $529,923$ are flagged as robust candidates, according to the quality flag introduced to take into account the \cite{croom2009} selection criteria. This catalogue will be made publicly available through the CDS VizieR facility. As discussed in the text, the catalogue requires to be filtered according to the specific needs of the user.

\section*{Acknowledgments}
The authors wish to thank the anonymous referee for all very useful comments and suggestions which greatly helped to clarify and to improve the readability of the paper.
The authors wish to acknowledge the financial support from the PRIN 2011 MIUR grant \textit{Cosmology with Euclid}.
MB acknowledges financial support from PRIN-INAF 2014 \textit{Glittering Kaleidoscopes in the sky, the multifaceted nature and role of galaxy clusters}.
We made use of Topcat tool developed within the Virtual Observatory, and the data mining infrastructure DAMEWARE.
This research has made use of the SDSS III DR10 and VizieR catalogue data access tools.

\label{lastpage}

\begin{thebibliography}{99}


\bibitem[\protect\citeauthoryear{Abraham et al.}{2012}]{abraham2012} Abraham S., Philip N.S., Kembhavi A., Wadadekar Y.G., \& Sinha R., 2012, MNRAS, 419, 80

\bibitem[\protect\citeauthoryear{Ahn et al.}{2014}]{ahn2013} Ahn C.P., et al. 2014, ApJS, 211, 2, 17

\bibitem[\protect\citeauthoryear{Antonucci}{1993}]{antonucci1993} Antonucci R., 1993, ARA\&A, 31, 473

\bibitem[\protect\citeauthoryear{Borne}{2010}]{borne2010} Borne K., 2010, Bulletin of the American Astronomical Society, 42, 578

\bibitem[\protect\citeauthoryear{Brescia et al.}{2012}]{brescia2012} Brescia M., Cavuoti S., Paolillo M., Longo G., Puzia T.H., 2012, MNRAS, 421, 1155

\bibitem[\protect\citeauthoryear{Brescia et al.}{2013}]{brescia2013} Brescia M., Cavuoti S., D'Abrusco R., Mercurio A., Longo G., 2013, ApJ, 772, 140

\bibitem[\protect\citeauthoryear{Brescia et al.}{2014}]{brescia2014} Brescia, M.; Cavuoti, S.; Longo, G. et al., 2014, PASP, 126, 942, 743-797

\bibitem[\protect\citeauthoryear{Byrd et al.}{1994}]{byrd1994} Byrd R.H., Nocedal J., Schnabel R.B., 1994, Mathematical Programming, 63, 4, pp. 129-156

\bibitem[\protect\citeauthoryear{Cavuoti et al.}{2015}]{cavuoti2015} Cavuoti S., Brescia M., De Stefano, V., Longo G., 2015, Experimental Astronomy, Springer, 39, 1, pp.45-71

\bibitem[\protect\citeauthoryear{Cavuoti et al.}{2012}]{cavuoti2012} Cavuoti S., Brescia M., Longo G., Mercurio A., 2012, A\&A, 546, 13

\bibitem[\protect\citeauthoryear{Cavuoti et al.}{2014a}]{cavuoti2014} Cavuoti S.; Brescia M.; D'Abrusco R.; Longo G. \& Paolillo M., 2014, MNRAS 437, 968

\bibitem[\protect\citeauthoryear{Cavuoti et al.}{2014b}]{cavuoti2014b} Cavuoti S.; Brescia M.; Longo G., 2014, proceedings of the IAU Symposium, Vol. 306, Cambridge University Press

\bibitem[\protect\citeauthoryear{Croom et al.}{2009}]{croom2009} Croom, S.~M., Richards, G.~T., Shanks, T., et al., 2009, MNRAS, 392, 19-44

\bibitem[\protect\citeauthoryear{D'Abrusco et al.}{2009}]{dabrusco2009} D'Abrusco R.; Longo G., Walton N.A., 2009, MNRAS, 396, pp. 223-262

\bibitem[\protect\citeauthoryear{Dawson et al.}{ 2013}]{boss} Dawson K. S., et al. 2013, AJ, 145, 10


\bibitem[\protect\citeauthoryear{Djorgovski et al.}{2012}]{george12} Djorgovski S.G., Mahabal A., Drake A., Graham M., \& Donalek C. 2012, in Astronomical Techniques, Software, and Data (ed. H. Bond), Vol.2 of Planets, Stars, and Stellar Systems (ser. ed. T. Oswalt), p. 223. Berlin:Springer Verlag

\bibitem[\protect\citeauthoryear{Eisenstein et al.}{2011}]{eisenstein2011} Eisenstein D.J., et al. 2011, AJ, 142, 72

\bibitem[\protect\citeauthoryear{Ge et al.}{2008}]{marvels}	Ge J., Mahadevan S., Lee B., Wan X., Zhao B., et al., 2008, in ASP Conference Series, Vol. 398, Edited by D. Fischer et al., p.449
	
\bibitem[\protect\citeauthoryear{Geisser}{1975}]{geisser1975} Geisser S., 1975, Journal of the American Statistical Association, 70 (350), 320-328


\bibitem[\protect\citeauthoryear{Grossberg}{1973}]{grossberg1973} Grossberg S., 1973,  Studies in Applied Mathematics, 52, 213, 1973

\bibitem[\protect\citeauthoryear{Kewley et al.}{2001}]{kewley2001} Kewley, L.J.; Dopita, M.A.; Sutherland, R.S.; Heisler, C.A.; Trevena, J.; 2001, The Astrophysical Journal, Volume 556, Issue 1, p. 121-140

\bibitem[\protect\citeauthoryear{Lamareille}{2010}]{lamareille2010} Lamareille, F., 2010, A\&A, Vol. 509, id. A53

\bibitem[\protect\citeauthoryear{Ledermann}{1982}]{ledermann1982} Ledermann, W., 1982, Handbook of Applicable Mathematics (New York:Wiley), Vol.6

\bibitem[\protect\citeauthoryear{Lintott et al.}{2008}]{lintott2008} Lintott C.J., Schawinski K., Slosar A., Land K., Bamford S., et al., 2008, MNRAS, 389, 1179

\bibitem[\protect\citeauthoryear{Majewski et al.}{2014}]{apogee} Majewski S.R., and SDSS-3/APOGEE Coll., 2014, AAS Meeting, 223

\bibitem[\protect\citeauthoryear{Messias et al.}{2010}]{messias2010} Messias H., Afonso J., Hopkins A., Mobasher B., Dominici T., Alexander D.M., 2010, ApJ, 719, 790

\bibitem[\protect\citeauthoryear{Palanque-Delabrouille et al.}{2013}]{palanque2013} Palanque-Delabrouille, N., et al., 2013, A\&A, 551, A29, 14 pp.

\bibitem[\protect\citeauthoryear{Pilbratt et al.}{2010}]{pilbratt2010} Pilbratt G.L., et al., 2010, A\&A, 518, L1

\bibitem[\protect\citeauthoryear{Richards et al. }{2002}]{richards2002}	Richards G.T., et al., 2002, AJ, 123, 2945

\bibitem[\protect\citeauthoryear{Richards et al.}{2004}]{richards2004} Richards G.T. et al., 2004, ApJS, 155, 257

\bibitem[\protect\citeauthoryear{Richards et al. }{2009}]{richards2009} Richards G.T., et al., 2009, ApJS, 180, 67-83

\bibitem[\protect\citeauthoryear{Rosenblatt}{1961}]{rosenblatt1961}Rosenblatt F., 1961, Principles of Neurodynamics: Perceptrons and the Theory of Brain Mechanisms. Spartan Books, Washington DC

\bibitem[\protect\citeauthoryear{Ross et al.}{2012}]{ross2012} Ross, N.~P., Myers, A.~D., Sheldon, E.~S., et al., 2012, ApJSS, 199, 3, 29 pp.

\bibitem[\protect\citeauthoryear{Schweitzer et al.}{2006}]{Schweizer2006} Schweitzer M., Lutz D., Sturm E., et al., 2006, ApJ, 649, 79


\bibitem[\protect\citeauthoryear{Sinha et al.}{2007}]{sinha2007} Sinha R.P., Philip N.S., Kembhavi A.K., \& Mahabal A.A., 2007, Highlights of Astronomy, 14, 609

\bibitem[\protect\citeauthoryear{Stoughton et al.}{2002}]{stoughton2002} Stoughton, C., et al., 2002, ApJ, 123, 485

\bibitem[\protect\citeauthoryear{Urry \& Padovani}{1995}]{padovani1995} Urry M.C., Padovani P., 1995, PASP, 107, 803


\bibitem[\protect\citeauthoryear{Winchatz \& Anderson}{2007}]{winchatz2007} Winchatz, B.~B., \& Anderson, S.~F., 2007, MNRAS, 374, 1506-1514

\bibitem[\protect\citeauthoryear{Yanny et al.}{2009}]{yanny2009} Yanny B., et al., 2009, AJ, 137, 4377

\bibitem[\protect\citeauthoryear{York et al.}{2000}]{york2000} York, D.G., et al., 2000, AJ, 120, 1579

\end{thebibliography}
\end{document}